\documentclass[11pt]{article}

\newcommand{\lb}{$\langle \:$}
\newcommand{\gb}{$\rangle \:$}

\newcommand{\alp}{$\alpha$\ }
\newcommand{\etal}{{\it et.al.}}
\newcommand{\degree}{$^{\circ}$ }

\newcommand{\kms}{km sec$^{-1}$\ }

\newcommand{\Msol}{$M_{\odot}$\ }

\newcommand{\Lsol}{$L_{\odot}$\ }

\usepackage{apjfonts}
\usepackage{float}
\usepackage{multicol}
\usepackage{epsfig}

\begin{document}

\twocolumn[
\null
\vspace{-1cm}

\begin{center}
\noindent

{\Large \bf Star Formation and Tidal Encounters with the Low Surface
Brightness Galaxy UGC 12695 and Companions}

\vspace{1cm}

K. O'Neil$^1$, M.A.W. Verheijen$^2$, S.S. McGaugh$^3$ \\

\end{center}

\vspace{1cm}

$^1$Arecibo Observatory, HC03 Box 53995, Arecibo, PR 00612, email:koneil@naic.edu \\
$^2$National Radio Astronomy Observatory, Box 0, Socorro, NM 87801, email:mverheij@nrao.edu \\
$^3$Department of Astronomy, University of Maryland, College Park, MD 20742, email:ssm@astro.umd.edu \\

\vspace{0.5cm}

\begin{flushright}

\parbox{13.3cm}{{\large\noindent{\bf\sf ABSTRACT-- }}
We present VLA H I observations of the low surface brightness galaxy UGC
12695 and its two companions, UGC 12687 and a newly discovered dwarf
galaxy 2333+1234.  UGC 12695 shows solid body rotation but has a very
lopsided morphology of the H I disk, with the majority of the H I lying
in the southern arm of the galaxy.  The H I column density distribution
of this very blue, LSB galaxy coincides in detail with its light
distribution.  Comparing the H I column density of UGC 12695 with the
empirical (but not well understood) value of $\Sigma_c$ = 10$^{21}$
atoms cm$^{-2}$ found in, i.e., Skillman's  1986 paper shows the star
formation to be a local affair, occurring only in those regions where
the column density is above this star formation threshold.  The low
surface brightness nature of this galaxy could thus be attributed to an
insufficient gas surface density, inhibiting star formation on a more
global scale.  Significantly, though, the  Toomre criterion places a
much lower critical density on the galaxy ($\sim$10$^{20}$ atoms
cm$^{-2}$), which is shown by the galaxy's low SFR to not be applicable.

Within a projected distance of 300 kpc/30 \kms\ of UGC 12695 lie two
companion galaxies -- UGC 12687, a high surface brightness barred spiral
galaxy, and 2333+1234, a dwarf galaxy discovered during this
investigation.  The close proximity of the three galaxies, combined with
UGC 12695's extremely blue color and regions of localized starburst and
UGC 12687's UV excess bring to mind mutually induced star formation
through tidal activity. } 
\end{flushright}  

\vspace{0.5cm}

\noindent
{\it Subject headings:} 
galaxies: individual(UGC 12695, UGC 12687, 2333+1234) 
-- galaxies: spiral
-- galaxies: evolution   
-- galaxies: interactions
-- galaxies: kinematics and dynamics
-- galaxies: structure

\null
\vspace{0.5cm}
]

\setcounter{footnote}{0}
\footnotetext{To be published in The Astronomical Journal, May 2000}

\newpage

\begin{figure*}[t]
\begin{center}
\epsfig{file=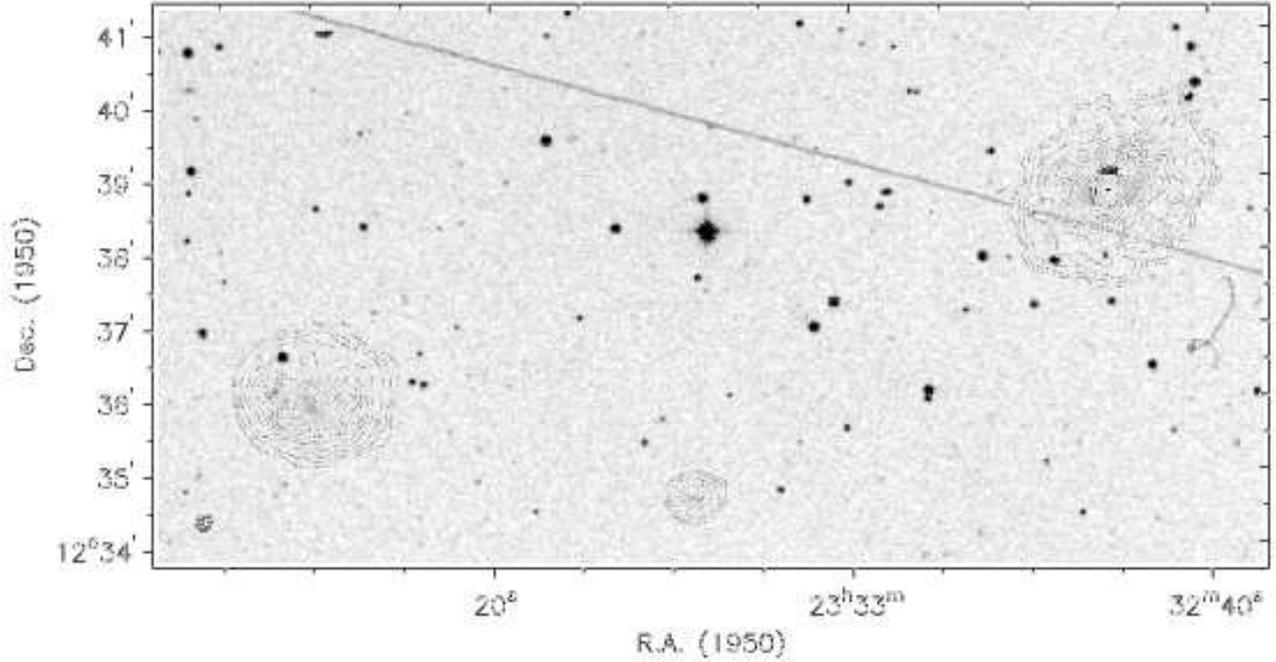,width=17cm}
\caption{H I contours of all three galaxies overlaid  on a POSS-II
image.}
\label{fig:Allposs2HI}
\end{center}
\end{figure*}

\section{Introduction}

Attempts to understand star formation in low surface brightness (LSB)
galaxies has resulted in a large number of theories being discarded and
few alternatives being offered.  As a result we have considerable
knowledge on what these enigmatic systems are not.  LSB galaxies are
{\em not}:

\begin{itemize}

\item simply the faded version of high surface brightness (HSB)
galaxies.  Although some red LSB galaxies have been found which may be
the end product of the faint blue galaxies, the majority of LSB galaxies
have very blue colors and low metallicities (i.e.  Ferguson \& McGaugh
1995; O'Neil, \etal 1997a; McGaugh 1994; Schombert, \etal 1990; De Blok
\& Van der Hulst 1998), arguing against any fading scenario. 

\item lacking the neutral hydrogen necessary to form stars, as many LSB
galaxies contain more than 10$^9$ \Msol of H I and LSB galaxies include
some of the highest M$_{HI}$/L$_B$ galaxies known (O'Neil, Bothun, \&
Schombert 1999). 

\item a completely new type of galaxy.  The transition from HSB to LSB
galaxies is smooth, with LSB galaxies covering the entire color and
morphological spectrum of HSB galaxies (i.e.  O'Neil, \etal 1997b;
Matthews \& Gallagher 1997)

\end{itemize}

UGC 12695 is a relatively nearby (z=0.021) low surface brightness galaxy
with an absolute blue magnitude of M$_B$=$-$18.9.  Previous studies of
UGC 12695 (McGaugh, 1994; O'Neil \etal, 1998) have shown it to be very
remarkable.  The galaxy is of an exceedingly transparent nature,
evidenced by the many background galaxies seen through its elusive disk,
and it contains a reasonably high gas fraction (M$_{HI}$/L$_B$ = 2.6
\Msol/\Lsol) while having a very low metallicity and an extremely blue
color for a galaxy ($U-I$ = $-0.2$) (Table 1). 

Because UGC 12695 was thought to be fairly isolated, with the nearest
galaxy (UGC 12687) lying more than 277 kpc away
(Figure~\ref{fig:Allposs2HI}), it provides a good opportunity for
studying star formation and evolution in LSB galaxies.  To this end, and
with the above points in mind, we undertook to observe UGC 12695 with
the Very Large Array (VLA) in the C configuration.  The results of these
observations are described in this paper, as follows: Section 2
describes the observations and data reduction; Section 3 examines the H
I morphology and kinematics of UGC 12695 and its companions --  UGC
12687, and 2333+1234; Section 4 looks at the dark and visible mass of
UGC 12695; Section 5 examines the  star formation potential of UGC
12695; Finally, section 6 examines the possibility of a recent tidal
encounter between the UGC galaxies. 

\begin{table}[ht]
\begin{center}
\caption{Global properties of UGC~12695 and UGC~12687.}
\begin{tabular}{lccc}
\noalign{\vspace{0.8mm}}
\hline
\hline
\noalign{\vspace{0.8mm}}
                    &            &     UGC~12695     &    UGC~12687         \\
\hline
\noalign{\vspace{0.8mm}}
Type          &            &            Sm$^1$ &              SBbc$^1$\\
V$_{hel}$           & km s$^{-1}$&           6186$^2$&               6150$^2$\\
M$_B$               & mag        &          -18.9$^3$&              -20.3$^1$\\
$B-V$               &            &          0.26$^3$ &               0.70$^4$\\
$\mu_B(0)$     &mag/arcsec$^{2}$&          23.8$^3$ &                 -     \\
r$_{25}$            & kpc        &          0.79$^1$ &               0.87$^1$\\
M$_{HI}$               & M$_\odot$  & 7.5$\times$10$^9$ $^2$& 1.2$\times$10$^{10}$ $^2$\\
M$_{HI}$/L$_B$ & M$_\odot$/L$_\odot$ &      2.62$^2$ &              1.18$^2$ \\
\noalign{\vspace{0.8mm}}
\hline
\multicolumn{4}{l}{$^1$De Vaucouleurs, \etal 1991}\\
\multicolumn{4}{l}{$^2$This paper}\\
\multicolumn{4}{l}{$^3$O'Neil, \etal 1998}\\
\multicolumn{4}{l}{$^4$Prugniel \& Heraudeau 1998}\\
\noalign{\vspace{0.8mm}}
\hline
\hline
\end{tabular}
\end{center}
\end{table}

\begin{table}[ht]
\caption{VLA observing parameters}
\begin{tabular}{lr}
\noalign{\vspace{0.8mm}}
\hline
\hline
\noalign{\vspace{0.8mm}}
Configuration                                               &     C-short \\
Correlator mode                                             &  2AC-normal \\
Total integration time \hfill                       (hours) &        15.5 \\
Dates of observation                                        &     15Jan99 \\
\multicolumn{2}{r}{16Jan99} \\
Field center,           $\alpha$(B1950)                     &    23:33:30 \\
\phantom{Field center, }$\delta$(B1950)                     &    12:35:53 \\
Central frequency \hfill                              (MHz) &     1391.64 \\
$V_{\rm hel}$ of central channel \hfill    (km$\:$s$^{-1}$) &        6170 \\
Primary beam FWHM \hfill                           (arcmin) &        32.4 \\
Synthesized beam ($\alpha$$\times$$\delta$) \hfill (arcsec) & 16.2$\times$14.1 \\
Bandwidth \hfill                                      (MHz) &      1.5625 \\
Number of channels                                          &         256 \\
Channel separation \hfill                  (km$\:$s$^{-1}$) &        1.31 \\
Velocity resolution \hfill                 (km$\:$s$^{-1}$) &        1.58 \\
rms noise in one channel \hfill                         (K) &        2.63 \\
K-mJy conversion,                                           &             \\
\phantom{K-}equiv. of 1mJy/beam \hfill                  (K) &        2.76 \\
\noalign{\vspace{0.8mm}}
\hline
\hline
\end{tabular}
\end{table}

\section{The Data -- Observations and Reduction}  

H I spectral line synthesis observations of UGC 12695 and its companions
were done in two runs with the VLA in its new C-short configuration and
are specified in Table~2.  The primary calibrator 3C48 was observed
three times per run and the secondary phase calibrator 2340+135 was
observed every 35 minutes. 

Calibration, flagging, concatenation and Fourier transformation of the
UV data was done with the AIPS package.  A robust R=0 weighting of the
UV data points was applied and the entire primary beam was imaged with a
512 x 512 map of 5 arcsecond pixels.  The dirty maps and corresponding
antenna patterns were exported into the GIPSY package which was used for
further data reduction and analysis as described below. 

The number of velocity channels was reduced by averaging adjacent pairs
of channel maps which resulted in a data cube of 127 nearly independent
channels, each 2.68 \kms\ wide.  The dirty maps were cleaned down to
half the rms noise level with channel dependent search areas using the
standard H\"ogbom algorithm.  The clean components were restored with a
Gaussian beam of FWHM 16.2$\times$14.1 arcsec at a position angle of
$-$54 degrees.  The data cube was then Hanning smoothed in velocity
which resulted in a velocity resolution of 5.3 \kms. 

No continuum emission was detected in the averaged line-free channels at
the positions of UGC 12695 and 2333+1234.  Due to a steep rotation
curve, the line emission at the center of UGC 12687 spans the entire
bandpass, leaving only 4 line-free continuum channels at the high
velocity end of the data cube while some line emission at the low
velocity edge of the bandpass is severely affected by the high noise
level.  No continuum emission could be detected in the line-free
channels at the position of the disk of UGC 12687.  Therefore, no
continuum map was subtracted to avoid the unnecessary addition of noise
to the channel maps and the nuclear continuum emission of UGC 12687 was
removed at a later stage. 

The areas of H I emission were isolated in each channel map and the
pixels outside these areas were set to zero.  Global H I profiles were
derived by measuring the total flux in the isolated areas, corrected for
primary beam attenuation.  In the case of UGC 12687, a 2.9 mJy baseline
was subtracted from the global profile. 

Integrated H I maps of the galaxies were constructed by summing the
primary beam corrected, isolated areas of H I emission.  At the position
of the nucleus of UGC 12687, 7 channels at the high velocity end of the
data cube are free from line emission and those were averaged to obtain
a map of the central continuum source.  A Gaussian beam was fitted to
this source, giving a primary beam corrected flux density of 2.9$\pm$0.2
mJy at the position $23^{\rm h}32^{\rm m}45^{\rm s}.4$ and $23^{\rm
d}32^\prime53^{\prime\prime}$ (B1950). Subsequently, this fitted
Gaussian was subtracted from the integrated H I map.

Velocity fields were constructed by fitting a single Gaussian to each
profile and rotation curves for UGC 12695 and UGC 12687 were derived by
fitting tilted rings of 11 arcsec width to their velocity fields. 

Optical images were taken from the Hubble Space Telescope Wide Field
Planetary Camera-2 (WFPC2) images of O'Neil, \etal (1998), the MDM 1.3m
McGraw Hill telescope (McGaugh, Schombert, \& Bothun 1995), and from the
Space Telescope Science Institute Digital Sky Survey.  The metallicity
studies of UGC 12695 are from McGaugh (1994). 

H$_0$ is 75 km s$^{-1}$ Mpc$^{-1}$ throughout this paper, and a 
Virgo-centric infall of 300 km s$^{-1}$ is assumed.  B1950 coordinates
are used throughout this paper.

\section{H I morphology and kinematics}

The following subsections contain detailed descriptions of the overall
properties of the neutral hydrogen gas in UGC 12695 and its companions
UGC 12687 and 2333+1234 which are illustrated in
Figures~\ref{fig:U12695}, \ref{fig:U12687}, \ref{fig:2333},
respectively.  The beam size is 16.2'' $\times$ 14.1'', or 6.4 kpc
$\times$ 5.6 kpc at 82 Mpc. 

\subsection{UGC 12695}

\begin{figure*}[p]
\epsfig{file=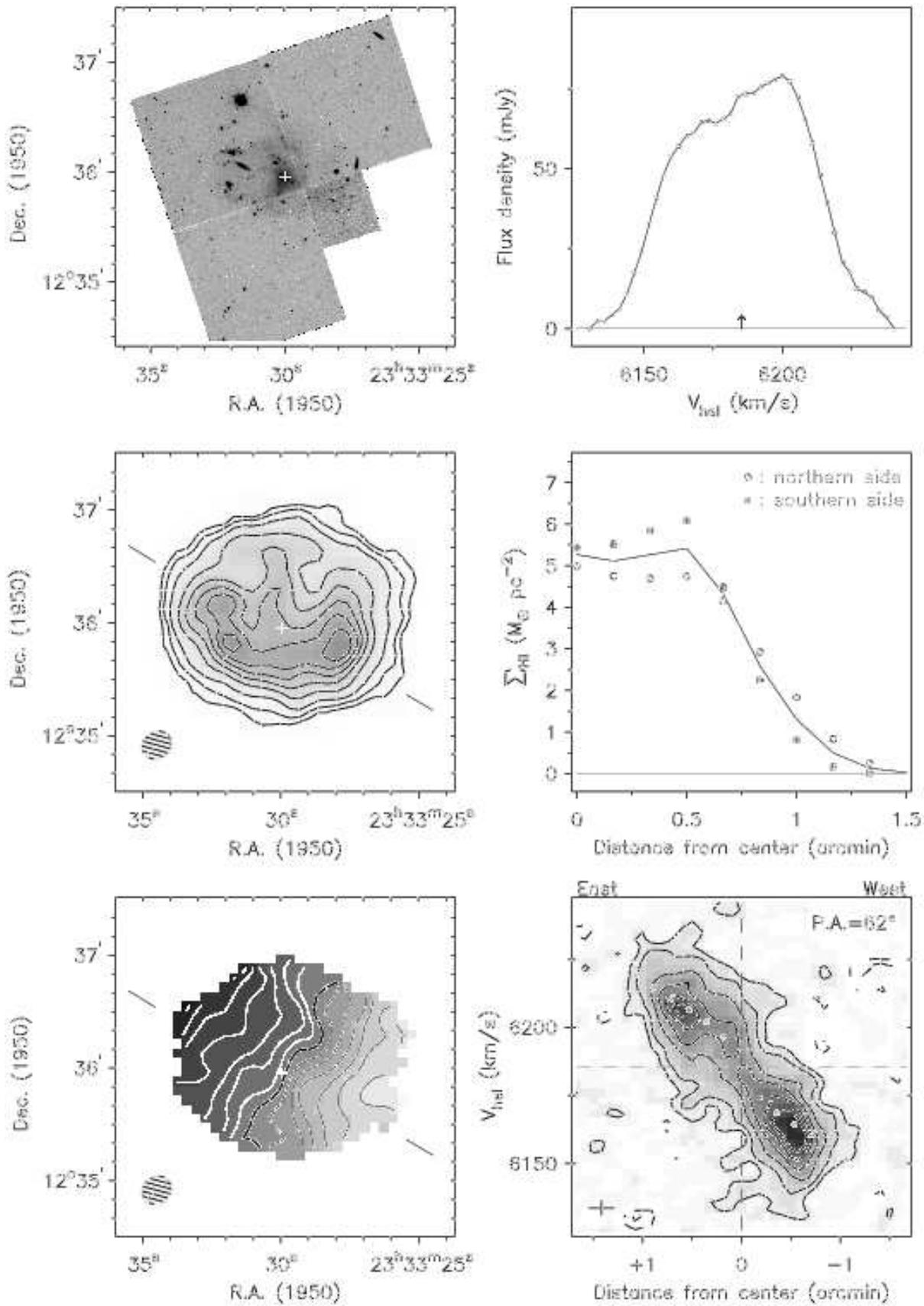,width=16cm}
\caption{UGC 12695. See section 3.1 for explanations.}
\label{fig:U12695}
\end{figure*}

Figure~\ref{fig:U12695} presents the data of UGC 12695.  The upper left
panel displays the HST WFPC2 F814W image of O'Neil, \etal\ (1998).  It
shows a relatively smooth triangular inner region and an irregular outer
disk dominated by several large star forming H$\alpha$ regions.  Several
background galaxies can be seen through the disk, evidencing its
extremely transparent nature.  The southern spiral arm  seems to be
sharply outlined while the northern arm is extremely diffuse.  Smoothing
the WFPC2 image to a 1$^{\prime\prime}$ resolution and fitting an
ellipse to the faintest isophotes indicates a position angle of
88$^\circ$, an inclination of 43$^\circ$ and a central position at
(23$^{\rm h}$33$^{\rm m}$30.4$^{\rm s}$,
12$^\circ$36$^\prime$1$^{\prime\prime}$).

The upper right panel shows the global H I profile obtained by measuring
the flux in the individual channel maps.  The width at the 20\% level of
the peak flux is 79.4 \kms\ and the width at the 50\% level of the peak
flux is 62.2 \kms.  The integrated flux density is 4.7 Jy km s$^{-1}$
which corresponds to a total H I mass of 7.5$\times$10$^9$ M$_\odot$ for
a distance of 82 Mpc (v=6186 km s$^{-1}$ (Table 1) and H$_0$=75 km
s$^{-1}$ Mpc$^{-1}$).  The shape of the profile suggests a global
lopsidedness of the H I distribution and or kinematics.  The vertical
arrow indicates the systemic velocity as derived from the H I velocity
field.   It should be noted that the H I profile of UGC 12695 was
previously determined both by Theureau, \etal\ (1998) using the
Nan\c{c}ay telescope and Schneider, \etal (1990) using the Arecibo
telescope.   Although both of the earlier observations match our
velocity widths,  the Nan\c{c}ay result list a 40\% smaller total flux. 
As our results match those of Schneider, \etal, we believe the data
differences to be the result of uncertain beam shapes and primary beam
corrections in the Nan\c{c}ay data.

The middle left panel presents the integrated H I column density map
with the size of the synthesized beam in the lower left corner.  This H
I map is at the same scale as the WFPC2 image above.  Contour levels are
drawn at 0.5, 1, 2, 4, 6, 8, 10 and 12$\times$10$^{20}$ atoms cm$^{-2}$.
 Overall, the neutral hydrogen distribution of UGC 12695 appears to
match the optical morphology quite well, including the fact that the H I
distribution is very lopsided with a high column density ridge running
through the southern part of the disk.  The cross corresponds to the
position of the cross in the WFPC2 image and indicates the central
optical concentration.  Fitting an ellipse to the lowest H I contours
indicates a position angle of 80$^\circ$, an inclination of 37$^\circ$
after a first order beam smearing correction, and puts the center of the
H I disk at (23$^{\rm h}$33$^{\rm m}$30.0$^{\rm s}$,
12$^\circ$36$^\prime$4$^{\prime\prime}$), 7 arcseconds ($<$ 1 beam
width) north of the central optical concentration. 

The middle right panel shows the radial H I column density distribution,
azimuthally averaged over the northern and southern sides separately.
Clearly, the H I surface density falls off more sharply at the southern
edge, going from 10 to 0.5 x 10$^{20}$ atoms cm$^{-2}$ within two beam
widths. 

The lower left panel shows the H I velocity field.  Apart from some
obvious wrinkles due to non-circular or streaming motions, the velocity
field is dominated by solid body rotation.  This makes it impossible to
determine the dynamical center and inclination from the velocity field
and therefore the optical center (cross) was adopted as the dynamical
center.  The thick line indicates the adopted systemic isovelocity
contour at 6185.7 \kms\ while the black contours indicate the
approaching side and the white contours the receding side of the galaxy.
 The isovelocity contour intervals are set at $\pm$n$\times$5 \kms. 

The lower right panel presents the position-velocity diagram along the
kinematic major axis.  Contours are drawn at -4, -2 (dashed), 2, 4, 8,
12, 16 and 20 times the rms noise level.  The vertical dashed line
corresponds to the position of the cross in the left panels, the
horizontal dashed line corresponds to the adopted systemic velocity. 
The cross in the lower left corner indicates the beam.  All profiles in
the vertical direction can be well described by single Gaussians.  No
double profiles are observed.  The solid points show the derived
rotation curve projected onto the position-velocity diagram.  The
rotation curve was derived by fitting full tilted rings to the velocity
field, effectively azimuthally averaging the wrinkles.  Consequently,
this azimuthally averaged rotation curve might deviate locally from the
position-velocity slice. 

The rotation curve of UGC 12695 is tabulated in Table~3.  Fitting a
single, galaxy wide ring to the entire velocity field gives a position
angle of the kinematic major axis of 62 degree.  The short thin lines
outside the velocity field indicate this average kinematic major axis. 
Note the significant difference of 18 degrees between the kinematic and
morphological position angles of the outer H I disk.  An inclination of
40$^\circ$ is adopted which is the average of the optical and H I
inclinations.  Given this rather face-on orientation of the disk, the
uncertainty in the position of the dynamical center and the obvious
deviations from circular motions, we estimate the uncertainties in the
rotation curve at some 20\%.

\subsection {UGC 12687}

\begin{figure*}[p]
\epsfig{file=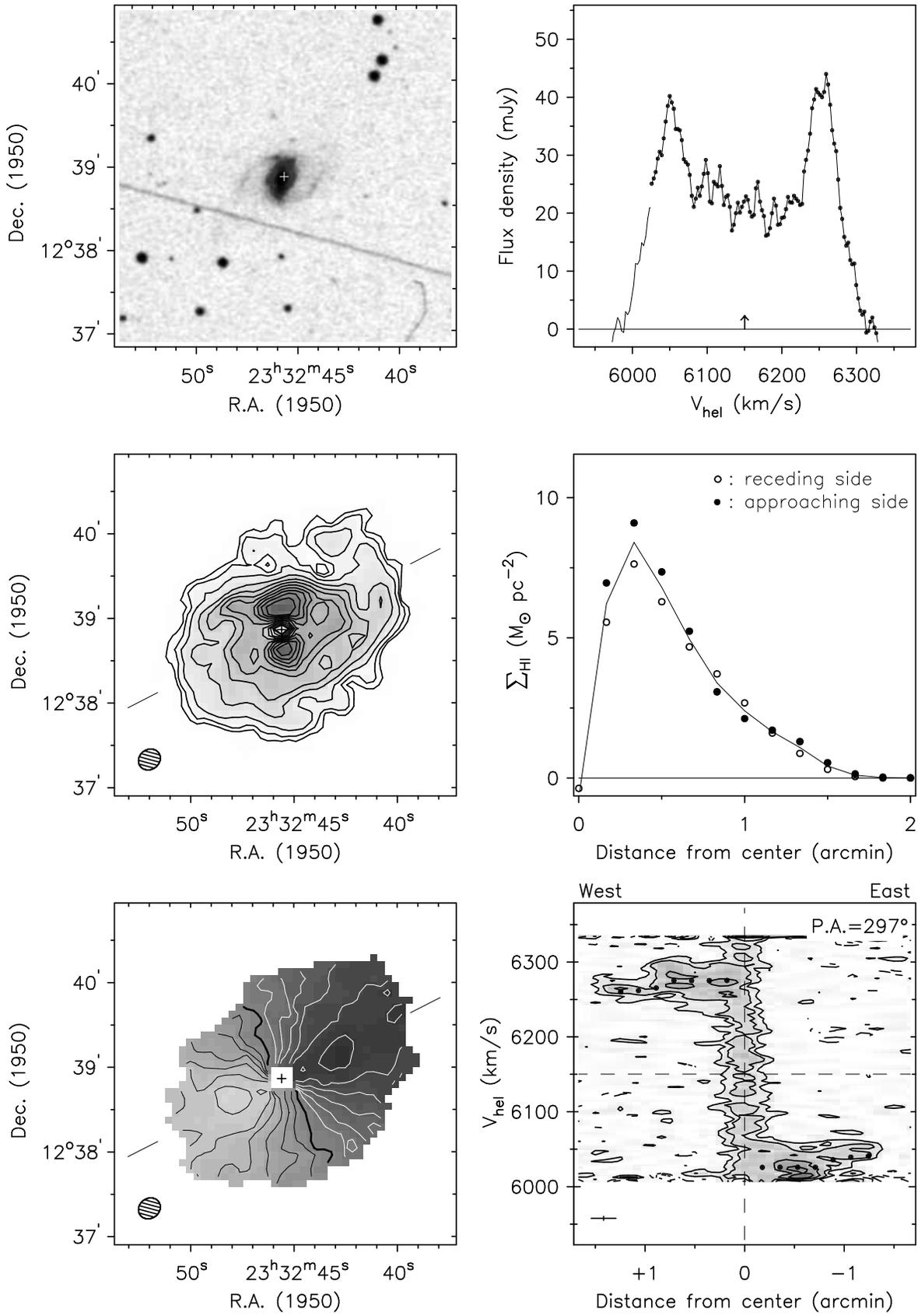,width=16cm}
\caption{UGC 12687. See section 3.2 for explanations.}
\label{fig:U12687}
\end{figure*}

\begin{figure*}[p]
\epsfig{file=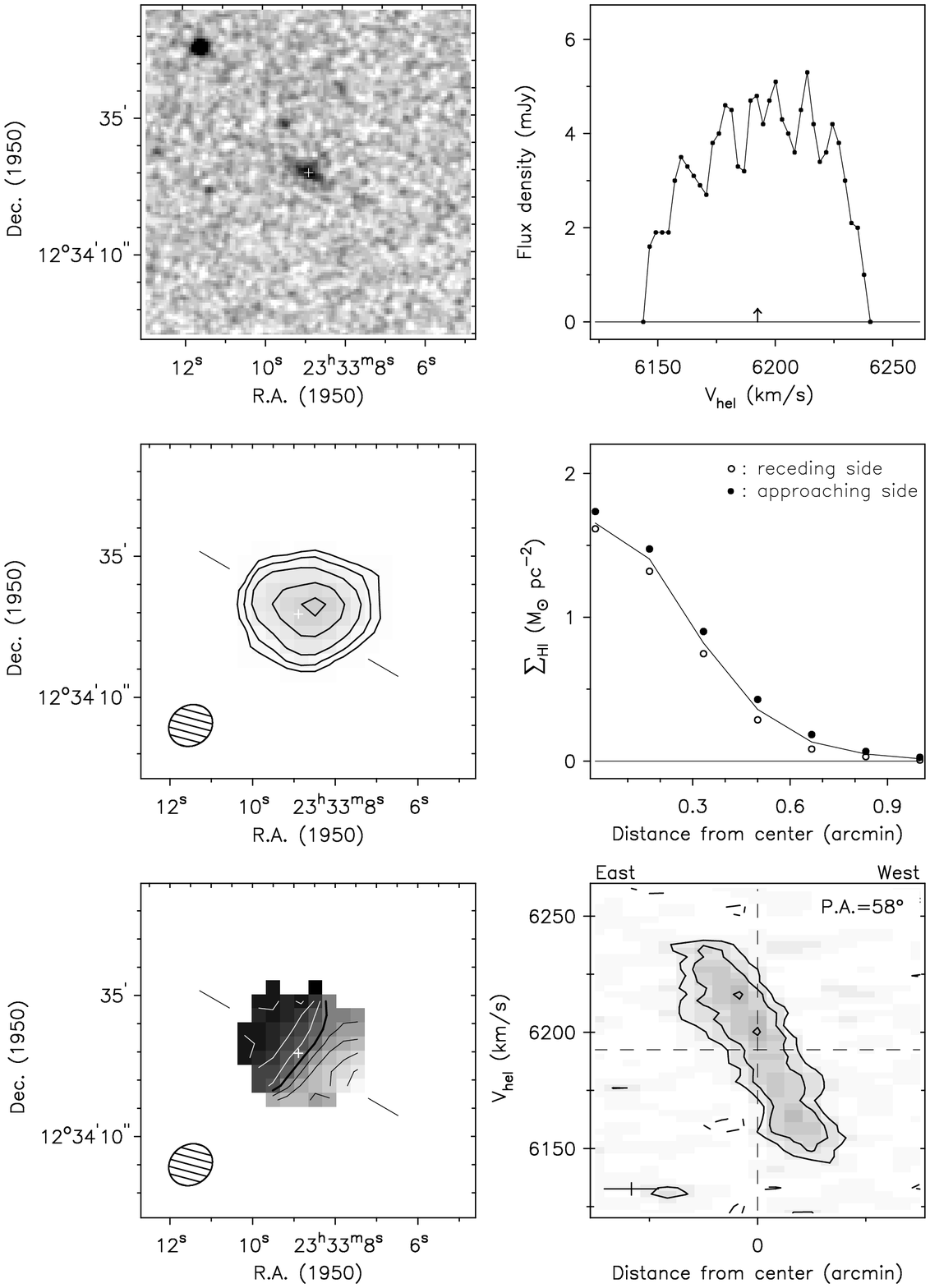,width=16cm}
\caption{2333+1234. See section 3.3 for explanations.}
\label{fig:2333}
\end{figure*}

The upper left panel of Figure~\ref{fig:U12687} shows the blue POSS-II
image of UGC 12687, a strongly barred two-armed spiral.  The bar
dynamics efficiently feeds gas to the nuclear region where a radio
continuum source with a peak flux of 4.0$\pm$0.6 mJy is found at 1.4 GHz
(Condon {\it et al}, 1998).  An ultra-violet excess has been reported by
Kazarian \& Kazarian (1985), suggesting a high level of star formation
activity.  Nevertheless, the B$-$V=0.70 color from Prugniel \& Heraudeau
(1998)  of UGC 12687 is considerably redder than that of UGC 12695. 

The upper right panel shows the global H I profile which displays the
classical double-horned shape.  Fluxes were measured in individual
channel maps including the central continuum source.  Afterwards, a
2.9~mJy/beam continuum baseline was subtracted from the global profile. 
Unfortunately, H I emission at the lower velocities is lost in the edge
of the passband.  To estimate total fluxes and line widths, the high
velocity edge was mirrored and, as an educated guess, put in place of
the missing low velocity side of the profile.  This technique gives an
integrated flux density of 7.5 Jy km s$^{-1}$ or a total H I mass of
1.2$\times$10$^{10}$ M$_\odot$.  The inferred line widths are 296.7
\kms\ at the 20\% level and 255.4 \kms\ at the 50\% level.   Like UGC
12695, UGC 12687 was imaged by Theureau, \etal\ (1998) with the
Nan\c{c}ay telescope, with similar results -- the velocity widths of the
 Nan\c{c}ay data matched ours well, but the total flux reported by 
Theureau, \etal\ was only 80\% of our result.  To check our data, we
obtained a 5 minute ON/OFF pair with the Arecibo telescope using the
L-narrow receiver.  The Arecibo data and our VLA data again matched to
within 5\% in total flux.

The middle left panel displays the integrated column density map of UGC
12687 constructed by adding the individual channel maps, including the
central continuum source which was removed by subtracting a 2.9 mJy/beam
central point source.  Contour levels are drawn at 0.5, 1, 2, 4, 6, 8,
10, 12, 14, 16 and 18$\times$10$^{20}$ atoms cm$^{-2}$.  The central
hole in the H I map might be due to a slight overestimation of the
continuum flux or might be caused by H I seen in absorption. 
Furthermore, the approaching south-eastern side of the galaxy is missing
some flux in the integrated H I map due to the bandpass effect mentioned
above.  Nevertheless, it is clear that the H I gas in UGC 12687 is
concentrated near the tips of the bar and to some extent along both
optically visible spiral arms.  Fitting an ellipse to the outer H I
contours gives an axis ratio of (b/a)=0.72 and a position angle of 129.8
degrees centered on (23$^{\rm h}$32$^{\rm m}$45.2$^{\rm s}$,
12$^\circ$38$^\prime$54$^{\prime\prime}$). 

The middle right panel shows the azimuthally averaged radial H I surface
density profiles of the receding and approaching sides separately. Note
that the approaching side misses some flux around a radius of 1
arcminute. 

The lower left panel shows the velocity field which suggests, at least
in projection, a declining rotation curve in the inner regions.  Fitting
tilted rings gives a dynamical center at (23$^{\rm h}$32$^{\rm
m}$45.4$^{\rm s}$, 12$^\circ$38$^\prime$52$^{\prime\prime}$), a systemic
velocity of 6150.2 \kms\ (thick line), an inclination of 43$^\circ$ and
a position angle of 297$^\circ$.  However, due to the strong bar,
non-circular motions are certainly present.  No significant warp could
be detected. The isovelocity contours are plotted at intervals of
$\pm$n$\times$20 \kms. The inferred rotation curve of UGC 12687 is
tabulated in Table~3.

The lower right panel shows the position-velocity diagram over the
entire observed bandwidth along the kinematic major axis.  The central
continuum source has not been removed.  Note how the low velocity gas is
lost in the edge of the bandpass as well as the limited number of line
free channels at the high velocity side. Also note the occasional double
profiles. 

\subsection{2333+1234}

In the VLA data cube, the H I emission of a tiny irregular dwarf  low
surface brightness galaxy was discovered.  Having discovered it first in
H I, we were then able to discern the galaxy as a barely visible smudge
on the POSS-II plate (left panel of Figure~\ref{fig:2333}).  Fitting an
ellipse to the faintest POSS-II isophotes yields a size of
17.5$\times$7.2 arcsec and a position angle of 58$^\circ$, centered on
(23$^{\rm h}$33$^{\rm m}$8.9$^{\rm s}$,
12$^\circ$34$^\prime$39$^{\prime\prime}$). 

The upper right panel shows the measured global H I profile with an
integrated flux of 0.33 Jy km s$^{-1}$ or a total H I mass of
5.2$\times$10$^8$ M$_\odot$. The line widths are 92 \kms\ at the 20\%
level and 75 \kms\ at the 50\% level.

The middle left panel shows the resolved integrated H I column density
map which seems to be slightly offset from the optical image.  H I
contours are plotted at 0.5, 1, 2, 4 and 6$\times$10$^{20}$ atoms
cm$^{-2}$

The middle right panel shows the barely resolved radial H I surface
density profile. No deconvolution attempt was made. 

The lower left panel shows the velocity field which clearly indicates a
velocity gradient along the optical major axis.  The optical center was
taken to be the dynamical center and a systemic velocity of 6192.5 \kms\
was inferred.  Isovelocity contours are plotted in steps of
$\pm$n$\times$10 \kms.  Obviously, trying to derive a rotation curve by
fitting tilted rings is futile. 

The lower right panel displays the position-velocity diagram through the
optical center along the kinematic major axis, however, and the sign of
solid body rotation is evident.

\section{Dark and Visible Matter in UGC 12695}

\begin{table}[t]
\begin{center}
\caption{Inclination corrected rotation curves of UGC~12695 and
UGC~12687.}
\begin{tabular}{ccccccc}
\noalign{\vspace{0.8mm}}
\hline
\hline
\noalign{\vspace{0.8mm}}
  &\multicolumn{3}{c}{UGC~12695} & \multicolumn{3}{c}{UGC~12687} \\
R & V$_{\rm rot}$ & incl. & P.A. & V$_{\rm rot}$ & incl. & P.A. \\
$\prime\prime$ & km/s & $\circ$ & $\circ$ & km/s & $\circ$ & $\circ$ \\
\hline
\noalign{\vspace{0.8mm}}
 10.7 & 17 & 40 & 62 & 195 & 43 & 297 \\
 21.3 & 26 & 40 & 62 & 195 & 43 & 297 \\
 32.0 & 32 & 40 & 62 & 195 & 43 & 297 \\
 42.7 & 38 & 40 & 62 & 195 & 43 & 297 \\
 53.3 & 45 & 40 & 62 & 179 & 43 & 297 \\
 64.0 & 52 & 40 & 62 & 174 & 43 & 297 \\
 74.7 &    &    &    & 171 & 43 & 297 \\
\noalign{\vspace{0.8mm}}
\hline
\hline
\end{tabular}
\end{center}
\end{table}

Previous studies of the dark and visible matter of LSB galaxies have
shown them to be extremely dark matter dominated with respect to
``normal'' HSB galaxies (i.e. Van Zee, \etal 1997; De Blok \& McGaugh
1997, 1998).  Thus, although the lack of any turn-over in UGC 12695's
rotation curve makes it clear that we have not come close to determining
the full gravitational potential of the galaxy, it is still a worthwhile
exercise to look at UGC 12695's total mass.

\begin{figure}[t]
\epsfig{file=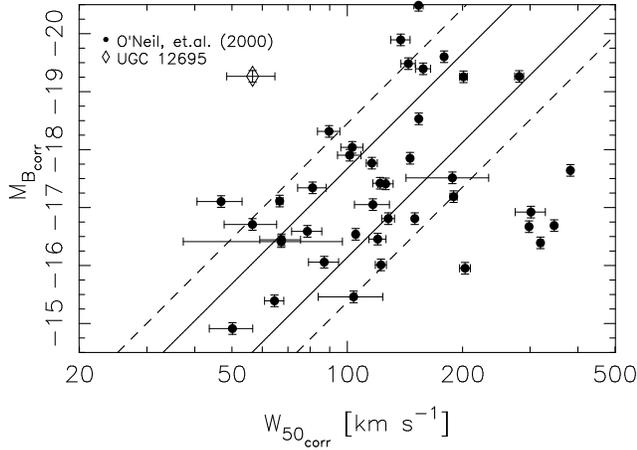,width=8.5cm}
\caption{LSB galaxy Tully-Fisher relation.  The 
lines are the 1$\sigma$ and 2$\sigma$ fits to the data of Zwaan, \etal
(1995), and the circles are recent observations of LSB galaxies in the    
Pegasus and Cancer groups from O'Neil, Bothun, \& Schombert (1999).  UGC   
12695 (using {\it i}  = 40\degree) is shown as a diamond.}
\label{fig:tf}
\end{figure}

Classic Newtonian mechanics states that the dynamical mass of a
rotating, gravitationally bound object  is simply
\[M_{dyn}\:=\:{{v^2\:R}\over{G\:\sin^2{\it i}}}\] where G is the
gravitational constant. Using the maximum known velocity of UGC 12695
(33/sin(40\degree) km s$^{-1}$ at r=64''), this gives a total dynamical
mass of 16 x 10$^{9}$\Msol, while  the determined H I flux gives a total
H I mass of M$_{HI}$ = 7.5 x 10$^{9}$\Msol.  Although at first glance
these numbers hardly seem remarkable, they imply a considerable absence
of dark matter  for a LSB galaxy.  Assuming a minimal disk scenario
(M$_*$/L$_B$ = 0), and letting all the gas in the galaxy be neutral
hydrogen  and helium (M$_{gas}$ = 1.47$\times$M$_{HI}$ =
11$\times$10$^9$\Msol), gives a dark-to-total mass ratio of only
M$_{DM}$/M$_{dyn}$ = 0.30.  Using somewhat more realistic numbers by
letting M$_*$/L$_B$ = 1, (a low-to-average LSB maximal disk value from
Van Zee, \etal\ 1997 \& De Blok \& McGaugh 1997) reduces the dark matter
contribution to only 12\% of the total dynamical mass of UGC 12695. 
(The luminosity value, L$_B$ = 2.86 $\times$ 10$^9$ \Msol, is derived
from the value given in O'Neil, \etal, 1998 which used integrated
aperatures. The error in L$_B$ is less than 1\%.) For comparison, the
average M$_{DM}$/M$_{dyn}$ values for LSB galaxies from De Blok \&
McGaugh is 0.6 for maximum disk scenarios, and for Van Zee, \etal\
(1997)  \lb~M$_{DM}$/M$_{dyn}$\gb = 0.7.  Additionally, if the stellar
mass-to-light ratio of UGC 12695 is increased to 1.7, a reasonable value
for both HSB and LSB galaxies, there is no need to invoke any dark
matter to explain the maximum {\it observed} rotational velocity at the
last measured point of UGC 12695's rotation curve.  It should be  noted
that we were not able to observe any turn-over in UGC 12695's rotation
curve. Thus, unlike the Van Zee, \etal\ and De Blok \& McGaugh samples
we are not determining the dynamical mass from the flat portion of the
rotation curve but instead from the still rising portion. As such, it is
extremely likely that dark matter will play a large role in UGC 12695's 
outer regions.

It should also be noted that the {\it observed} velocity width of UGC
12695 causes it to fall  well off the standard Tully-Fisher relation,
lying approximately 2.5 magnitudes (3$\sigma$) above the LSB galaxy line
defined by Zwaan, \etal\ (1995) (Figure~\ref{fig:tf}).  This may be the
consequence of the apparent lack of dark matter in the observed portion
of UGC 12695.  On the other hand,  it is quite likely that there is
significant dark matter outside the observed radius (else the rotation
curve would show  some turn over), and thus we are merely viewing a
lower limit of the galaxy's rotational velocity. The uncertainty in UGC
12695's inclination (see the next section) also makes the current
location of UGC 12695 on the Tully-Fisher relation suspect and if the
inclination is less than 40\degree, UGC 12695 could move onto (or even
to the right of) the Tully-Fisher relation of Zwaan, \etal.

\section{Star Formation in UGC 12695}

\begin{figure*}[t] 
\epsfig{file=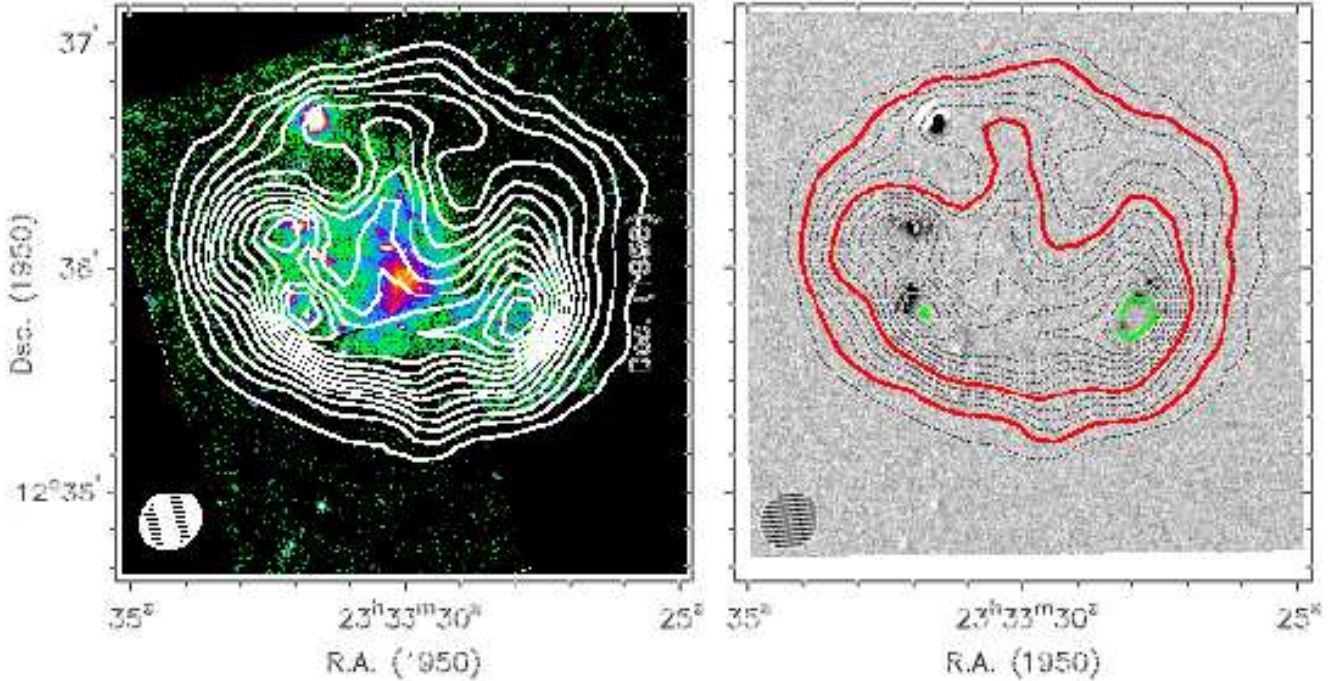,width=17.5cm}
\caption{Left panel: HI contours of U12695 overlaid  on the HST WFPC2
F814W image.  Right panel: HI contours of U12695 overlaid on the
H$\alpha$ image. The red contours indicate the critical HI column
density of 1.6$\times$10$^{20}$ and 6.0$\times$10$^{20}$ cm$^{-2}$ above
which star formation is expected based on Toomre's criterion (for {\it
i} = 40\degree \& 10\degree, respectively), while the green contours are
the 10$^{21}$ cm$^{-2}$ level.}
\label{fig:U12695optHI}
\end{figure*}  

It was demonstrated by Toomre (1964) that a thin, {\it collisionless}
stellar disk in circular motion becomes unstable if the surface mass
density exceeds a critical value of \[\Sigma_c\:=\:\alpha\:{{\kappa
\sigma}\over{3.36 G}}\] where $\Sigma_c$ is the critical density,
$\sigma$ is the velocity dispersion, \alp is a dimensionless constant
near 1, and $\kappa$ is the epicyclic frequency of the gas, also written
as \[\kappa\:=\:1.41\:{V\over R}\:\left(1\:+\:{R \over
V}{{dV}\over{dR}}\right)^{1/2}.\]  Cowie (1981) showed that this
criterion is also applicable to instabilities in a gaseous disk if
embedded in a more massive stellar disk.  Kennicutt (1989) determined an
empirical value for \alp of about 2/3.  Typical HSB galaxies exceed this
critical surface density and form stars throughout most of their stellar
disks.  

As an LSB galaxy which appears to be in the midst of considerable but
localized star formation, UGC 12695 is an ideal case on which to test
this star formation threshold theory.  Before this can be done, though,
$\kappa$ and $\sigma$ must be determined.  From the rotation curve of
UGC 12695 it is apparent that its near solid body rotation makes
determining $\kappa$ relatively easy.  Approximating the rotation curve
as pure solid body with an inclination corrected amplitude of 57 \kms\ 
at a radius of 22 kpc yields $\kappa$=5.2 \kms\ kpc$^{-1}$ (using the
fact that for a gas disk in pure solid body rotation,
${{dV}\over{dR}}={V\over R}$  and $\kappa={{2V}\over{R}}$).  Due to beam
smearing the velocity dispersion is hard to measure from the data and a
canonical dispersion of $\sim$8 \kms\ is assumed as an average estimate
(8 \kms\ is also observed in several highly resolved face-on gas disks).
 This leads to a critical surface mass density of
$\Sigma_c\:=\:4.0\times10^{-3}$ kg m$^{-2}$.  Taking a 32\% helium mass
fraction into account, this corresponds to a critical H I column density
of 1.6$\times$10$^{20}$ atoms cm$^{-2}$ (i.e. between the 1st and 2nd
contours in Figure~\ref{fig:U12695optHI}) above which star formation is
to be expected.  This implies that everywhere throughout the disk of UGC
12695 star formation should occur. 

\begin{figure*}[t]
\epsfig{file=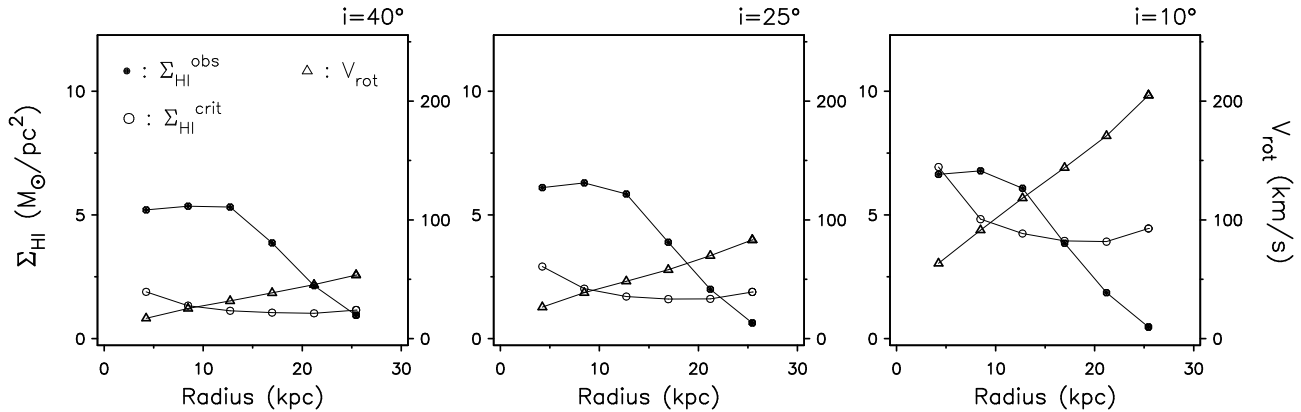,width=17.5cm}
\caption{The azimuthally averaged observed gas 
density, Toomre star formation threshold, and rotation curves of UGC
12695 for  an assumed galaxy inclination of 40\degree, 25\degree, and
10\degree (left to right).}
\label{fig:azav}
\end{figure*}

However, we only observe star formation in a limited number of localized
regions near the very peaks of the H I column density distribution where
it reaches levels of 1$\times$10$^{21}$ atoms cm$^{-2}$.  This is
illustrated in Figure~\ref{fig:U12695optHI} which shows in the left
panel the HI column density map overlaid on a false-color WFPC2 F814W
image and in the right panel the same H I contours overlaid on a
greyscale MDM-1.3m H$\alpha$ image. The lower red contour indicates the
critical column density of 1.6$\times$10$^{20}$ atoms/cm$^2$. Obviously,
the theoretically derived and empirically adjusted critical surface
density is clearly not applicable to the low metallicity, irregular gas
disk of UGC 12695.  

One of the more curious aspects of the Kennicutt-Cowie-Toomre star
formation criterion is that it successfully works at all, considering
the number of physical processes which affect the value of $\Sigma_c$. 
For example, disks are not infinitely thin but have a certain thickness
which could increase or decrease the column density thresholds and alter
the radial instabilities.  That is,  if the volume gas density is
significantly different than the surface gas density of UGC 12695, a
volume-density dependent Schmidt law would be more appropriate than the
Kennicutt/Cowie/Toomre star formation criterion used above (i.e.
Ferguson, \etal\ 1998).  Additionally, there is energy dissipation,
magnetic field lines, etc. which should also affect $\Sigma_c$ (i.e.
Hunter, Elmegreen, \& Hunter 1998).  Thus it is not surprising that UGC
12695, and in fact many LSB galaxies, do not adhere to the Toomre
criterion (i.e. Van Zee, \etal\ 1997; Van der Hulst, \etal\ 1993).

What is interesting is that UGC 12695, like many LSB and dwarf galaxies,
forms stars only where the local H I column density exceeds 10$^{21}$
atoms cm$^{-2}$.  In fact, Skillman (1986) pointed out that the actually
observed {\em local} H I column density threshold for star formation, at
a resolution of 500 pc, is about 1$\times$10$^{21}$ atoms cm$^{-2}$ and
roughly 5$\times$10$^{21}$ atoms cm$^{-2}$ for star formation events of
the order of 30 Doradus.  This local H I column density threshold
appears to be in better agreement with the observations of UGC 12695
(i.e. note the green contours in Figure~\ref{fig:U12695optHI}), although
the beam size makes a detailed analysis impossible. (The H$\alpha$ data
of McGaugh 1994 is not photometric, making determination of UGC 12695's
H II luminosity difficult.  It should be noted, though, that attempts to
detect faint diffuse H-$\alpha$ regions have not been  successful,
making it unlikely that any widespread component of faint star-forming
regions has been missed.)

Unlike our sample, a previous study by Van der Hulst, \etal\ (1993)
found their sample of low surface galaxies to be generally consistent
with  the Kennicutt-Toomre criterion for star formation.  Perhaps the
most important difference between this study of UGC 12695 and the Van
der Hulst, \etal\ results is that Van Der Hulst, \etal\ used azimuthally
averaged radial H I surface density profiles.  Figure~\ref{fig:azav}
shows the results of applying Van der Hulst, \etal's method to UGC 12695
for a variety of possible inclinations (see below). As can be seen, even
by ignoring the extremely asymmetric nature of UGC 12695, only the most
extreme case ({\it i}=10\degree) does UGC 12695 come close to falling
below the critical density for star formation anywhere but in the
outermost isophotes.  This sort of study, though, disallows for any
analysis of the local star forming potential of UGC 12695 while hiding
the exceptionally asymmetric nature of galaxy.

\begin{figure*}[p]
\hspace{-1cm}
\epsfig{file=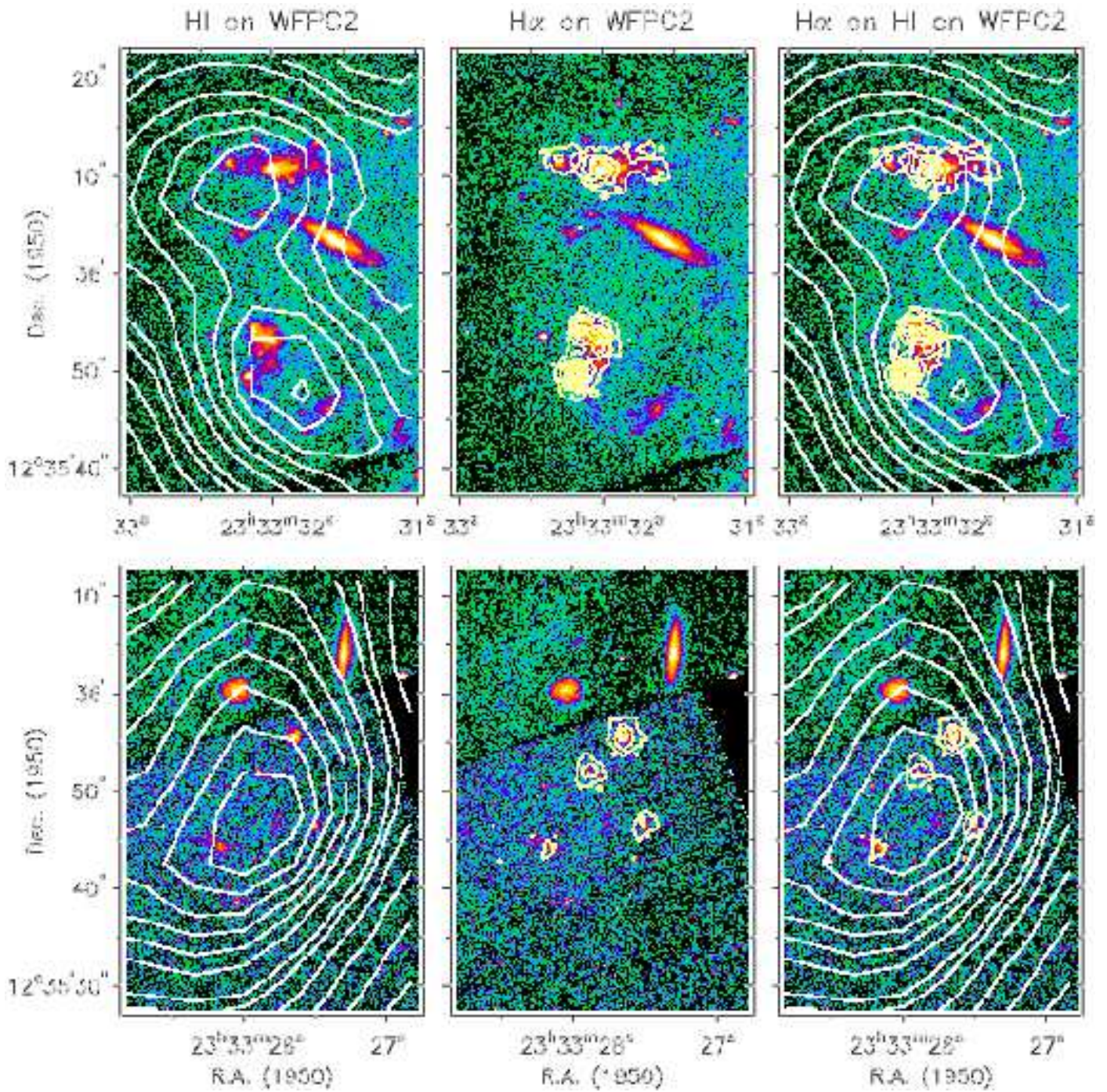,width=18.5cm}
\caption{Zooming in on the main star formation regions in in UGC~12695.
The upper row shows the star formation complexes in the eastern side of
the  galaxy. Those in the western side are shown in the bottom row.  The
white contours are from the VLA H I map while the yellow  contours are
from the MDM-1.3m H$\alpha$ image.}
\label{fig:U12695SFs}
\end{figure*}

At this point, it is important to consider the  uncertainties involved
in calculating $\Sigma_C$.  Most notably, we should take another look at
UGC 12695's assumed inclination. It is certainly possible that UGC
12695's shape truly is circular, thus validating the inclination value
used in the previous calculations (40\degree).  If, however, UGC 12695
has recently tidally interacted with UGC 12687, as discussed below, the
perceived inclination may be overestimated in that UGC 12695 may have
been distorted (and `flattened') by the interaction (e.g. see Figure 2
of Mihos, \etal, 1997). In this case the true inclination of UGC 12695
may be considerably less than we have assumed, thereby raising the 
value of $\Sigma_C$.  As an example, if UGC 12695's true inclination is
10\degree, the critical density  will increase to
$\Sigma_C\:=\:6\times10^{20}$ atoms cm$^{-2}$.  In this case, although
the critical density and the density at which star formation is observed
still would not precisely coincide, they would lie considerably closer 
together (i.e. the higher red contour in Figure~\ref{fig:U12695optHI}). 
If, in addition to the above correction to {\it i}, our estimate of
$\kappa$ is off by a factor of 60\% (3$\sigma$) due to inclination
uncertainties and  the rotation curve shape, the critical density would
readily be raised to 10$^{21}$ atoms cm$^{-2}$, the observed local H I
column density threshold for star formation of Skillman (1986). Of
course, if the inclination correction is off in the other direction, 
and {\it i}=50\degree, $\Sigma_C$ would be reduced even more, raising
again the question of why UGC 12695 is LSB.

It is noteworthy to point out that the three local peaks in the neutral
hydrogen of UGC 12695 lie near, but not on top, the primary star
formation regions of the galaxy, as defined by the H-\alp image of
McGaugh (1994). This is illustrated in Figure~\ref{fig:U12695SFs} which
displays in the left panels the white VLA H I contours on the color
WFPC2 F814W image, in the middle panels the yellow MDM-1.3m H$\alpha$
contours on the WFPC2 image and in the right panels the combined H I and
H$\alpha$ contours on the WFPC2 image.

Figure~\ref{fig:U12695SFs} shows that there seems to be a clear offset
between the highest peaks in the H I column density at 1.2 x 10$^{21}$
cm$^{-2}$ and the location of the primary H$\alpha$ complexes.  The
largest star clusters seem to surround the regions with the highest HI
column densities.  However, the relatively poor spatial resolution of
the current H I observations is insufficient to draw any further
conclusions on the relation between the H I peaks and the H$\alpha$
regions.
 
The colors of those regions, as provided by the WFPC2 images, also put
the star-formation peaks away from the H I peaks, with the left two H I
peaks having F300W $-$ F814W = $-$0.06 and $-$2.82, versus $-$3.27 and
$-$3.14 for the corresponding H-\alp peaks (top and bottom,
respectively).  (These colors roughly correspond to U $-$ I colors of
1.42, $-$1.34, $-$1.79.  and $-$1.66, respectively (i.e.  O'Neil, \etal
1998).) The third H I peak, at the bottom right of
Figure~\ref{fig:U12695optHI}, lies in the extremely noisy PC chip of the
WFPC2 image, making the determination of colors in that region extremely
difficult.  Thus the neutral hydrogen is behaving as expected -- as star
formation occurs the surrounding gas is ionized, shifting the peak in
the neutral hydrogen distribution to the edge of the star forming
regions.

\section{Are UGC~12695 and UGC~12687 Ti-dally Interacting?}

The close proximity of UGC 12695 and UGC 12687 in redshift space, the
lopsided morphology of UGC 12695 and its slightly skewed kinematics, 
immediately brings to mind the possibility of a tidal interaction
(Figure~\ref{fig:Allposs2HI}).  Additionally, the presence of 2333+1234
lying between the two galaxies suggests it may have been formed as a
tidal remnant.  (Of course, 2333+1234 may simply be a naturally
occurring representative of the faint-end of the luminosity function.) 

In 1997 Mihos, McGaugh, \& De Blok argued that LSB and HSB galaxies of
the same total mass are equally susceptible to local disk instabilities
but that LSB galaxies are far less responsive to global instabilities
than their HSB counterparts.  This difference is mainly due to the
stabilizing nature of the relatively more massive dark matter halo in
which the LSB disk is embedded.  To test their hypothesis, they modeled
a strong prograde tidal encounter between an LSB and an HSB disk galaxy
of similar mass.  After the encounter the HSB galaxy exhibited two
definitive spiral arms, a central inflow of gas and an oval central
region.  Presumably, the HSB system was in the midst of, or had recently
undergone, a large burst of star formation in its core.  Being more
stable than its HSB counterpart, the LSB galaxy displayed a milder, yet
significant response.  Although the encounter strongly perturbed the LSB
galaxy, it did not result in a central gas inflow.  However, it did
induce long-lived spiral arms, an overall lopsided distortion of the
galaxy, and possibly localized compressions and instabilities in the
disk. 

The observed morphologies of UGC 12695 and UGC 12687 show a striking
resemblance to the numerical simulations of Mihos \etal\ at the time
stamp T=36 (see their Figure~2).  Their HSB system (UGC 12687 in our
case) shows a strong bar from which two well defined spiral arms emerge.
 UGC~12687's observed morphology is in close agreement with their
results while its central continuum emission and UV-excess indicate a
considerable nuclear star formation activity, hinting at well developed
bar kinematics efficient in fueling H I to the central region.  In the
case of the LSB galaxy, the numerical simulations display a sharp
stellar edge on one side of the disk and a more diffuse gradient on the
other side while a highly variable structure in the mass surface density
hints at strong local instabilities.  Observationally, UGC 12695's
extremely blue colors, highly asymmetric gas and star distribution, and
regions of intense local star formation also match the model predictions
extremely well.  In fact, without relying on some sort of external
trigger the observed morphology and color of UGC 12695 is extremely
difficult to explain. 

In spite of the above assertions, a number of arguments against any
major tidal encounter between these two galaxies must be considered. 
The first, and perhaps most obvious of these is the apparently settled
kinematics of both UGC 12695 and UGC 12687.  At first glance it would
seem that if the two galaxies have interacted recently enough for the
tidally-induced star formation to be at, or near, its peak the galaxies
would still exhibit highly agitated kinematics.  A study by V\'azquez
and Scalo (1989), though, has shown that starbursts do not typically
occur during the gas compression stage but in fact occur well after the
gas has re-established.  In other words, the V\'azquez and Scalo model
suggests that disks can have tidally induced star formation well after
the gas has kinematically re-settled. 

A second argument which could be put forward against the idea of the two
galaxies having recently undergone a tidal interaction is simply this --
if UGC 12695 is experiencing a burst of localized star formation due to
a recent tidal encounter, should it not be experiencing a corresponding
rise in central surface brightness? A recent paper by O'Neil, Bothun, \&
Schombert (1998) tested this idea through modeling a wide variety of LSB
galaxies experiencing localized starbursts.  Their results were quite
definitive -- if a galaxy forms as a LSB galaxy, due to a high angular
momentum giving rise to a low gas surface density etc., it will remain a
LSB galaxy barring any major encounter catastrophe. Thus it is quite
believable that UGC 12695 could be undergoing significant localized star
formation activity and yet not be undergoing any significant change in
its global surface brightness. 

The final argument against UGC 12695 and UGC 12687 having undergone a
significant tidal interaction in the recent past comes from examining
the smoothed data cube.  Not a trace of extended H I gas above a minimal
detectable column density of 2$\times$10$^{19}$ atoms cm$^{-2}$
(3$\sigma$) can be found besides the rotating gas disks of the three
identified galaxies.  This leads to the conclusion that no major tidal
tails were ever formed in any past interaction between the two systems.

\section{Conclusion}
 
UGC 12695 is an intriguing low surface brightness galaxy of a very
transparent nature, having an extremely blue color, a highly asymmetric
appearance and very localized bursts of star formation near the peaks in
the H I column density distribution.

Many of the properties of both UGC 12687 and UGC 12695 can be explained
as being induced by such a tidal interaction, including the bar of UGC
12687 and its central radio continuum emission and UV excess as well as
the lopsided appearance of UGC 12695 and the offset between its
morphological and kinematic major axes.  Furthermore, the localized
bursts of star formation in UGC 12695 could very well be induced by such
an interaction, giving rise to local instabilities in the LSB disk as
demonstrated by Mihos \etal (1997). 

It is likely that UGC 12695 could have been living a fairly quiescent
existence, its low surface gas density keeping its star formation rate
quite low, and just now it is experiencing a period of localized but
vigorous star formation triggered by a mild tidal interaction which
might lead to a major morphological transition. 

Within all this, though, it is easy to overlook one important fact. 
Although many of the properties of UGC 12695 and UGC 12687 can readily
be explained through an ongoing tidal encounter, the two galaxies are
still fundamentally distinct.  UGC 12695 is not simply a fainter, or
more `stretched-out', or more quickly rotating version of UGC 12687. 
Were any of these the case the behavior of the two galaxies after the
tidal encounter would be similar, and UGC 12695 would have experienced a
central gas inflow with the majority of its star formation now occurring
not in the outlying regions (as it is), but in the galaxy's core.  Thus
the fundamental question of why UGC 12695 is an LSB galaxy, and UGC
12687 is not remains unanswered.

\section{Acknowledgments}

The Very Large Array is a facility of the National Radio Astronomy
Observatory, a facility of the National Science Foundation operated
under cooperative agreement by Associated Universities, Inc.  The
Digitized Sky Surveys were produced at the Space Telescope Science
Institute under U.S.  Government grant NAG W-2166.  The images of these
surveys are based on photographic data obtained using the Oschin Schmidt
Telescope on Palomar Mountain and the UK Schmidt Telescope.  The plates
were processed into the present compressed digital form with the
permission of these institutions.

\section*{References}

\noindent
 Condon, \etal, 1998, AJ, 115, 1693\\

\vspace{-3.5mm}
\noindent
 Cowie, L., 1981, ApJ, 245, 66\\

\vspace{-3.5mm}
\noindent
 De Blok, W.J.G. \& Van der Hulst, J.M., 1998, A\&A, 336, 49\\

\vspace{-3.5mm}
\noindent
 De Blok, W.J.G. \& McGaugh, S., 1998, ApJ, 499, 41\\

\vspace{-3.5mm}
\noindent
 De Blok, W.J.G. \& McGaugh, S., 1997, MNRAS, 290, 533\\

\vspace{-3.5mm}
\noindent
 Ferguson, H., \& McGaugh, S., 1995, ApJ, 440, 470\\

\vspace{-3.5mm}
\noindent
  Ferguson, A. \etal, 1998, ApJ, 506, L19\\

\vspace{-3.5mm}
\noindent
 Kazarian, M.A. \& Kazarian, E.S., 1985, Afz, 22, 431\\

\vspace{-3.5mm}
\noindent
 Kennicutt, R., 1989, ApJ, 344, 685\\ 

\vspace{-3.5mm}
\noindent
 Hunter, D., Elmegreen, B. \& Baker, A., 1998, ApJ, 493, 595\\

\vspace{-3.5mm}
\noindent
 Matthews, L., \& Gallagher, J., 1997, AJ, 114, 1899\\

\vspace{-3.5mm}
\noindent
 McGaugh, S., 1994, ApJ, 426, 135\\

\vspace{-3.5mm}
\noindent
 Mihos, C., McGaugh, S., \& De Blok, W.J.G, 1997, ApJ, 477L, 79\\

\vspace{-3.5mm}
\noindent
 O'Neil, K, Bothun, G., \& Schombert, J., 1999, AJ, in press\\

\vspace{-3.5mm}
\noindent
 O'Neil, K., Bothun, G., \& Schombert, 1998, AJ, 116, 2776\\

\vspace{-3.5mm}
\noindent
 O'Neil, K., \etal, 1998, AJ, 116, 657\\

\vspace{-3.5mm}
\noindent
 O'Neil, K., Bothun, G., \& Cornell M., 1997b, AJ, 113, 1212\\ 

\vspace{-3.5mm}
\noindent
 O'Neil, K., \etal, 1997a, AJ, 113, 1212\\

\vspace{-3.5mm}
\noindent
 Prugniel,  P. \& Heraudeau, P., 1998, A\&AS, 128, 299\\

\vspace{-3.5mm}
\noindent
 Schombert, J. \etal, 1990, AJ, 100, 1533\\

\vspace{-3.5mm}
\noindent
 Skillman, E.D., 1986, in "{\it Star Formation in Galaxies}", C.J. Lonsdale (ed), NASA Conference Publication 2466, 263\\

\vspace{-3.5mm}
\noindent
 Theureau, G., \etal, 1998, A\&AS, 130, 333\\

\vspace{-3.5mm}
\noindent
 Schneider, S., \etal, 1990, ApJS, 72, 245\\

\vspace{-3.5mm}
\noindent
 Toomre, A., 1964, ApJ, 139, 1217\\

\vspace{-3.5mm}
\noindent
 De Vaucouleurs, G. \etal, 1991, {\it The Third Reference Catalog of Bright Galaxies}, Springer-Verlag: New York\\

\vspace{-3.5mm}
\noindent
 Van der Hulst, J.M., \etal, 1993, AJ, 106, 548\\

\vspace{-3.5mm}
\noindent
 Van Zee, L., \etal, 1997, AJ, 113, 1618\\

\vspace{-3.5mm}
\noindent
 V\'azquez, E. \& Scalo, J.M., 1989, ApJ, 343, 644\\

\vspace{-3.5mm}
\noindent
 Zwaan, M., Van der Hulst, J., De Blok, W., \& McGaugh, S.S., 1995, MNRAS, 273, L35\\

\end{document}